\documentclass[12pt]{amsart}

\usepackage{amssymb, verbatim}

\usepackage{color,epsfig}

\usepackage{enumerate}

\date{\today}

\theoremstyle{definition}

\theoremstyle{remark}

\title[Some light on the OPERA most celebrated experiment]{Some light on ``Measurement of the neutrino velocity with the OPERA detector in the CNGS beam"}
\author[Area]{I. Area}
\author[Prado]{X. Prado}

\address[Area]{Departamento de Matem\'{a}tica Aplicada II, E.E. Telecomunicaci\'{o}n, Universidade de Vigo, 36310-Vigo, Spain.}\email[Area]{area@uvigo.es}
\address[Prado]{Departamento de F\'{\i}sica e Qu\'{\i}mica, I.E.S. Pedra da Auga, R\'ua Feliciano Barrera, s/n 36860 Ponteareas, Galiza, Spain.}\email[Prado]{yogote@edu.xunta.es}

\begin{document}

\begin{abstract}
The recent publication in ArXiv of ``Measurement of the neutrino velocity with the OPERA detector in the CNGS beam" has attracted many interest due to the possible theoretical or phenomenological interpretation of the results. A subtle broadening effect on the size of the neutrino pulse appearing in the OPERA measurements is analyzed in this paper to show that the results are in agreement with ``classical" theory of relativity.
\end{abstract}
\maketitle

\section{Introduction}

During several years many experiments done in the CERN have focused special interest not only for the scientific community but for journalism and therefore, the impact factor of many of the works done is extremely high. Among them, the manuscript \cite{OPERA} entitled ``Measurement of the neutrino velocity with the OPERA detector in the CNGS beam", appeared in arXiv on 22nd september 2011, has probably attained the highest social repercussion in last years. Note that a search in google with the words speed and neutrino has more than 2 million answers (only in english and at the time of writing this manuscript) with headlines as ``Tiny Neutrinos May Have Broken Cosmic Speed Limit" or ``Particles break light-speed limit".

At the time the snow ball started to run, Area contacted with Prado, who recently defended his Ph.D. on ``didactics of theory of relativity"  sending an email with just a few words in the subject: `What will you do now?" and a link to the manuscript \cite{OPERA} (28/09). After that, Prado's first reaction was ``it seems they are measuring the phase speed of the neutrino wave function instead of the neutrino speed itself". Some interchange of emails and ideas led us to write a few lines on this subject in order to give some light to this experiment. In the meanwhile, Shi-Yuan Li \cite{YUAN} has published a manuscript in arXiv in the same direction as proposed here (phase speed measuring). Also Mbelek \cite{Mbelek} suggested that the dispersion effect should be taken into account in the interpretation of OPERA experiment. Some other results have been appeared in arXiv, in both directions: trying to analyze the results and to prove that they are in agreement with theory of relativity or even assuming the results of the OPERA experiment in order to get new properties. We would like to mention here that already in the 1980's theoretical work has shown how to include supraluminal velocities in the framework of theory of relativity \cite{Recami1,Recami2,Recami3}.

After reading the OPERA manuscript \cite{OPERA} in search of a phase velocity  ---and therefore supraluminal---, we found a subtle broadening effect on the size of the neutrino pulse appearing in the OPERA measurements. A review of \cite{OPERA} under the light of this idea has proven that it is possible to reconcile these measurements with the theory of relativity.

The structure of the paper is as follows: in Section \ref{Sec:2}, we reread the OPERA experiment \cite{OPERA}. The subtle broadening effect is analyzed in Section \ref{Section:sbe}. In Section \ref{Sec:3} we  show that this effect is enough to reconcile the experimental results with the theory of relativity.

\section{The OPERA experiment}\label{Sec:2}

As described by the authors \cite{OPERA}, the CNGS beam is produced by accelerating protons to $400$ GeV/c with the CERN Super Proton Synchrotron (SPS). These protons are ejected with a kicker magnet towards a $2$ m long graphite neutrino production target in two extractions, each lasting $10.5$ $\mu$s and separated by $50$ ms. These pulses of neutrinos were produced at the CERN facilities in Prevesino with an average energy of $17$ GeV. The neutrinos were focused in the direction of a straight line that stretches from CERN to the detection systems in Gran Sasso, in Central Italy. The CERN CNGS beam had in both locations an angle of inclination with the horizontal of $3$ degrees due to the curvature of the Earth: in the central part of its trajectory, the neutrinos were moving $11$ km under the surface. The time of a neutrino interaction is defined as that of the earliest hit in the detector. The shapes of both extractions are also given in the manuscript. The main objective of the experiment consisted in the detection of neutrino oscillations in direct appearance mode, and it was later redesigned to allow determination for the neutrino velocity. In doing so, the timing system and the distance measurements were upgraded to reduce the systematic uncertainties. This resulted in a value for the distance of $731.2780$ km, with an uncertainty of $20$ cm.
In the experiment \cite{OPERA}, the authors measured also an early arrival time of neutrinos with respect to the one computed by assuming the speed of light in vacuum, with a value of $60.7$ ns (with a total uncertainty of $14$ ns). This fact is considered an anomaly by the authors, who quantified its value as a difference of the neutrino velocity with respect to the speed of light:
\[
\frac{v-c}{c} = 2.48 \times 10^{-5},
\]
with a total uncertainty of  $0.6 \times 10^{-5}$.

From these data it is straightforward to estimate the following approximate values:
\begin{enumerate}
\item Time of flight of the neutrinos from CERN to Gran Sasso: $2.44$ ms.
\item Surplus velocity of the neutrinos with respect to light $(v-c)$:  $7.4$ km/s.
\end{enumerate}

It might be concluded that the OPERA experiment has measured a supraluminal speed for a material particle, thus violating what people commonly consider one of the basic statements of Einstein's theory of relativity:  ``it is impossible to measure velocities that are faster than the speed of light in vacuum". But supraluminal velocities are not impossible in nature, nor are they banned from the theory of relativity; on the contrary, they fit perfectly within the theory of relativity. To show that, we should first notice that the relativistic statement really says that there is an absolute speed in the nature of space-time that is not possible to surpass by matter or energy. But other velocities not corresponding to the previous effects can be greater than $c$ or even infinite, and the Lorentz transformation applies to both kind of velocities, approaching them to the same limit from opposite sides of space-time in a symmetrical way that does not allow them to surpass it, in the same way that one cannot reach to the other side of a mirror by getting more and more closer to it (only Alice did!).

Quantum theory incorporates supraluminal velocities from the beginning, when it associates a probability wave to any particle. The velocities of the particle and of its associated wave are mutually inverse, so that when a particle is at rest its associated wave has an infinite velocity, and when the particle is in movement with a velocity $V_{g}$ (group velocity), its associated wave has a corresponding velocity $V_{f}$ (phase velocity). In natural units (where the speed of light $c$ is equal to $1$) the relation between both velocities reads as follows:  
\begin{equation}
V_{f} =\frac{1}{V_{g}}.
\end{equation}
The fact that the phase velocity is infinite when the particle is at rest can be interpreted as an infinite velocity for the collapse of the quantum wave when we measure the position of a particle (for example, when a neutrino is detected). If the collapse were not simultaneous, then we could at the same time make a measure for the same particle in two different parts of its wave packet and have some probability to find the particle in both places. We would have created two particles where only one existed, an impossible violation of fundamental laws of nature.

\section{The subtle broadening effect}\label{Section:sbe}

As already mentioned, our starting point was looking for the presence of some effect related to phase velocities in order to explain the supraluminal velocities measured in \cite{OPERA}. Surprisingly we found that these velocities could be explained taking into account a subtle broadening effect on the time-shape of the neutrino pulses that were measured at Gran Sasso. In \cite[Figure 9]{OPERA} it is represented the summed proton waveforms of the OPERA events corresponding to the two SPS extractions for the 2009, 2010 and 2011 data samples. It is evident that the width of both pulses is greater than the corresponding width of the original pulse at CERN which is $10.5 \mu$s. This subtle difference is about $0.5 \mu$s and $0.3 \mu$s in the first and second extraction, respectively. Furthermore, in \cite[Figure 12]{OPERA} it is shown a zoom of the leading and trailing edges of the measured neutrino interaction time distributions for the two SPS extractions. These figures show two different slopes which explain the broadening as an uncertainty of the pulse width, which values are again $0.5 \mu$s and $0.3 \mu$s, respectively. These coincidences suggest to analyze them as real effects and therefore to explore the consequences of these facts.

We will show how a neutrino pulse broadening of this magnitude can explain why there were measured supraluminal velocities and at the same time the neutrino velocity remained lesser than the speed of light. This broadening of the pulse adds a suplementary velocity to the group velocity of the pulse itself (which, according to theory of relativity, is lesser than the speed of light $c$). In this way it is possible to get an experimental measure of neutrino speed greater than $c$. This measure combines both: group velocities and phase velocities. 

Let us quantify this fact by using the smaller value of $0.3 \mu$s (in the other case of $0.5 \mu$s the results would show group velocities to be even more small that $c$). This extra time affects both extremes of the pulse, and we are interested in knowing the broadening speed of the starting part, which is the half of this total value. Taking into account this advanced arrival time of $0.15 \mu$s for the leading edge of the pulse due to its broadening (while traveling almost at the speed of light in their $731.3$ km long trip during the measured time of $2.44$ ms) this gives an advance distance of about $45$m. Using the classical composition of velocities, this means an additional velocity for the front part of the pulse of $45$ m $/$ $2.44$ ms $\approx$ $18.4$ km/s. Note that in the OPERA measurement, the experimental value for the speed is just $7.4$ km/s faster than light speed.

The possibility to increase the group velocity in more than $18.4$ km/s due to the pulse broadening provides a justification for the appearance of these ``supraluminal" $7.4$ km/s as being a totally relativistic and quantum effect.

\section{The speed of neutrinos considering the broadening of the neutrino pulse}\label{Sec:3}

As we have shown previously, it is in agreement with the data that the neutrino pulses do not have a constant size; their size gets instead greater at a rate that makes them larger when arriving at Gran Sasso than when they were produced at CERN, only $2.44$ milliseconds before. This effect is known as the beam dispersion of the quantum wave packet, and is a common fact of particle beams. The dispersion is related to the group and phase velocities of the quantum packet.

We would like to emphasize that the neutrino velocity has to be identified with the centre of the pulse, corresponding also to its probability peak. If we consider a reference frame that is at rest with the neutrinos, the broadening of every pulse will be the same in opposite directions, due to space isotropy. For this reason  the centre of the pulse will remain also at rest. We can therefore identify the speed of the individual neutrinos with that of the centre of the pulse, and this will also be true for any other reference frame.
We will use the expression ``Broadening speed", $B_{s}$  , for the velocity at which the front edge of the pulse is advancing in relation to its centre. We have found at the experimental data a plausible value for $B_{s}$ of $18.4$ km/s.
In a reference frame that is at rest with Earth and both laboratories (CERN and Gran Sasso), a measurement of the arrival time for the front edge of the pulses will show an experimental velocity that we will call $V_{ex}$. The velocity of the neutrinos, $V_{n}$ , as was made at their proper reference frame, has to be identified with the velocity of the centre of the pulse.  If we add the broadening speed $B_{s}$ to the velocity of the pulse centre $V_{n}$, we get the velocity for the front part of the pulse. This is the velocity that has been carefully measured at the OPERA experiment, $V_{ex}$.

Thus, using the experimental value for $V_{ex}$ and the plausible value for $B_{s}$, we can get an estimation for the neutrino velocity, $V_{n}$:
\[
V_{n} = V_{ex} - B_{s} = c + 7.4 \text{ km}/\text{s} - 18.4 \text{ km}/\text{s} = c - 11 \text{ km}/\text{s}.
\]

Taking into account the broadening speed it is therefore possible to justify a neutrino velocity that is not supraluminal even when the experiment gets a supraluminal value. Thus, the OPERA experiment is in agreement with the theory of relativity.

\end{document}